\documentclass[12pt]{article}
\usepackage{graphicx}
\usepackage{amsmath}
\usepackage{amssymb}
\usepackage{hyperref}
\usepackage{cite}
\usepackage{datetime}
\usepackage[dvipsnames]{xcolor}
\usepackage{overpic}
\usepackage{ogonek}

\renewenvironment{thebibliography}[1]{
  \begin{oldthebibliography}{#1}
    \setlength{\itemsep}{3pt}
    \setlength{\parskip}{0pt}
                                     }
{
  \end{oldthebibliography}
}

\newcommand{\peii}{$\pi_{e2}$}
\newcommand{\peiig}{$\pi_{e2\gamma}$}

\newcommand{\VmA}{$V$$-$$A$}  %%%% was {$V-A$}
\newcommand{\PR}{Phys.\ Rev.}
\newcommand{\PRL}{Phys.\ Rev.\ Lett.}

\newcommand{\NIM}{Nucl.\ Instrum.\ Meth.}

\renewcommand{\today}{\the\day\ \monthname\ \the\year}

\makeatletter

\renewcommand\section{\@startsection {section}{1}{\z@}%
                                   {-3.5ex \@plus -1ex \@minus -.2ex}%
                                   {2.3ex \@plus.2ex}%
                                   {\normalfont\large\bfseries\boldmath}}
\renewcommand{\@seccntformat}[1]{\csname the#1\endcsname.\quad}
\makeatother

\newdimen\arrayruleHwidth
\makeatletter
\def\Hline{\noalign{\ifnum0=`}\fi\hrule \@height \arrayruleHwidth
    \futurelet \@tempa\@xhline}
\makeatother
\newcommand\pubnumber{CIPANP2018-Pocanic}
\newcommand\pubdate{\today}

\def\support{\footnote{Work supported by US National Science Foundation
    (award PHY-1614839), the Paul Scherrer Institute, and the Russian
    Foundation for Basic Research.}}

\def\Title#1{\begin{center} {\Large #1 } \end{center}}
\def\Author#1{\begin{center}{ \sc #1} \end{center}}
\def\Address#1{\begin{center}{ \it #1} \end{center}}

\newcommand\pubblock{\rightline{\begin{tabular}{l} \pubnumber\\
         \pubdate  \end{tabular}}}
\newenvironment{Abstract}{\begin{quotation}  }{\end{quotation}}
\newenvironment{Presented}{\begin{quotation} \begin{center} 
         \textsc{presented at}\end{center}\bigskip 
      \begin{center} \begin{normalsize}}{\end{normalsize} 
       \end{center} \end{quotation}}

\def\beq{\begin{equation}}
\def\eeq#1{\label{#1}\end{equation}}
\def\eeqn{\end{equation}}

\def\beqa{\begin{eqnarray}}
\def\eeqa#1{\label{#1}\end{eqnarray}}
\def\eeqan{\end{eqnarray}}

\let\bar=\overbar

\def\Dslash{\not{\hbox{\kern-4pt $D$}}}
\def\dslash{\not{\hbox{\kern-2pt $\del$}}}

\def\msb{{\bar{\ssstyle M \kern -1pt S}}}

\begin{document}

\begin{titlepage}
\pubblock

\vfill
\Title{PEN experiment: a precise test of lepton universality}
\vfill
\Author{C.J.~Glaser,$^a$ D.~Po\v{c}ani\'c,$^{a\S}$
          L.P.~Alonzi,$^a$
          V.A.~Baranov,$^b$
          W.~Bertl,$^c$
          M.~Bychkov,$^a$
          Yu.M.~Bystritsky,$^b$
          E.~Frle\v{z},$^a$
          V.A.~Kalinnikov,$^b$
          N.V.~Khomutov,$^b$
          A.S.~Korenchenko,$^b$
          S.M.~Korenchenko,$^b$
          M.~Korolija,$^d$
          T.~Kozlowski,$^e$
          N.P.~Kravchuk,$^b$
          N.A.~Kuchinsky,$^b$
          M.C.~Lehman,$^a$
          E.~Munyangabe,$^a$
          D.~Mzhavia,$^{bf}$
          A.~Palladino,$^{ac}$
          P.~Robmann,$^{g}$
          A.M.~Rozhdestvensky,$^b$
%  U.~Straumann,$^{g}$
          R.T.~Smith,$^{a}$
          I.~Supek,$^d$
          P.~Tru\"ol,$^g$
%  Z.~Tsamalaidze,$^f$
          A.~van~der~Schaaf,$^g$
          E.P.~Velicheva,$^b$
          M.G.~Vitz,$^a$
          V.P.~Volnykh$^b$ \quad
           (The~PEN~Collaboration\support)
}
\Address{
  {$^a$}University of Virginia, Charlottesville, VA 22904-4714, USA\\ 
  {$^b$}Joint Institute for Nuclear Research, Dubna, Moscow Region, Russia\\
  {$^c$}Paul Scherrer Institute, Villigen-W\"urenlingen AG, Switzerland\\
  {$^d$}Rudjer Bo\v{s}kovi\'c Institute, Zagreb, Croatia\\
  {$^e$}Instytut Problem\'ow J\k{a}drowych im.\ Andrzeja
                 So{\l}tana, % PL-05-400 
                 \'Swierk, Poland\\
  {$^f$}Institute for High Energy Physics, Tbilisi State
                 University, % GUS-380086 
                 Tbilisi, Georgia \\
  {$^g$}Physik-Institut, Universit\"at Z\"urich, % CH-8057
                 Z\"urich, Switzerland \\[6pt]
  {$^{\S}$}Speaker and corresponding author, e-mail: {pocanic@virginia.edu}
}
\vfill
\begin{Abstract}
  With few open channels and uncomplicated theoretical description,
  charged pion decays are uniquely sensitive to certain standard model
  (SM) symmetries, the universality of weak fermion couplings, and to
  aspects of pion structure and chiral dynamics.  We review the current
  knowledge of the pion electronic decay $\pi^+ \to e^+\nu_e(\gamma)$,
  or $\pi_{e2(\gamma)}$, and the resulting limits on non-SM processes.
  Focusing on the PEN experiment at the Paul Scherrer Institute (PSI),
  Switzerland, we examine the prospects for further improvement in the
  near term.
\end{Abstract}
\vfill
\begin{Presented}
CIPANP 2018: 29 May\,--\,3 June 2018, Palm Springs, CA
\end{Presented}
\vfill
\end{titlepage}
\def\thefootnote{\fnsymbol{footnote}}
\setcounter{footnote}{0}

\section{Introduction: pion electronic decay, $\pi^+\to e^+\nu_e$}

Charged pion decays have provided important early insight into the
\VmA\ nature of the weak interaction following the failure of initial
searches to observe the direct electronic decay ($\pi\to e\nu$, or
\peii).  This led to a low branching fraction prediction of $\sim
1.3\times 10^{-4}$\cite{Fey58} as a consequence of the helicity
suppression of the right-handed state of the electron, even before the
decay's discovery\cite{Faz59}.  Further, predicted radiative
corrections for the \peii\ decay \cite{Ber58,Kin59} received quick and
decisive experimental confirmation \cite{And60,DiC64}, establishing the
process as an important theory testing ground.

Pion decays have more recently been described with extraordinary
theoretical precision.  Thanks to the underlying symmetries and
associated conservation laws, the more complicated, and thus more
uncertain, hadronic processes are suppressed.  Should measurement
results approach or reach the precision level of their theoretical
description, pion decays offer a uniquely clean testing ground for
lepton and quark couplings.  A statistically significant deviation from
the standard model expectations would indicate presence of processes or
interactions not included in the SM, affecting $\pi$ decays through loop
diagrams.

Of particular interest is the $\pi^- \to \ell\bar{\nu}_\ell$ (or,
$\pi^+\to \bar{\ell}\nu_\ell$) decay which connects a pseudoscalar $0^-$
state (the pion) to the $0^+$ vacuum.  At the tree level, the ratio of
the $\pi \to e\bar{\nu}$ to $\pi \to \mu\bar{\nu}$ decay widths is given
by \cite{Fey58,Bry82}
\begin{equation}
    R_{e/\mu,0}^\pi \equiv \frac{\Gamma(\pi \to  e\bar{\nu})}
          {\Gamma(\pi \to  \mu\bar{\nu})}
       = \frac{m_e^2}{m_\mu^2}\cdot
        \frac{(m_\pi^2-m_e^2)^2}{(m_\pi^2-m_\mu^2)^2}
      \simeq 1.283 \times 10^{-4}\,.  \label{eq:pi_e2_tree}
\end{equation}
The first factor in the above expression, the ratio of squared lepton
masses for the two decays, comes from the helicity suppression by the
\VmA\ lepton weak couplings to the $W$ boson.  If, instead, the decay
could proceed directly through the pseudoscalar current, the ratio
$R_{e/\mu}^\pi$ would reduce to the second, phase-space factor, or
approximately 5.5.  A more complete treatment of the process includes
$\delta R_{e/\mu}^\pi$, the radiative and loop corrections, and the
possibility of lepton universality (LU) violation, i.e., that $g_e$ and
$g_\mu$, the electron and muon couplings to the $W$, respectively, may
not be equal:
\begin{equation}
    R_{e/\mu}^\pi \equiv \frac{\Gamma(\pi \to  e\bar{\nu}(\gamma))}
          {\Gamma(\pi \to  \mu\bar{\nu}(\gamma))}
      = \frac{g_e^2}{g_\mu^2}\frac{m_e^2}{m_\mu^2}
        \frac{(m_\pi^2-m_e^2)^2}{(m_\pi^2-m_\mu^2)^2}
           \left(1+\delta R_{e/\mu}^\pi \right)\,,
    \label{eq:pi_e2_general}
\end{equation}
where the ``$(\gamma)$'' indicates that radiative decays are fully
included in the branching fractions.  Steady improvements of the SM
description of the $\pi_{e2}$ decay have reached the precision level of
8 parts in $10^5$: $R_{e/\mu}^{\pi,\,\text{SM}} = 1.2352(1) \times
10^{-4}$ \cite{Mar93,Fin96,Cir07}.  Comparison with equation
(\ref{eq:pi_e2_tree}) indicates that the radiative and loop corrections
amount to almost 4\% of $R_{e/\mu}^{\pi}$.  The current experimental
precision lags behind the above theoretical uncertainties by a factor of
$\sim$\,23: $R_{e/\mu}^{\pi,\,\text{exp}} = 1.2327(23) \times 10^{-4}$,
dominated by measurements from TRIUMF and PSI
\cite{Bri92a,Bri92b,Cza93,Agu15}.

Because of the large helicity suppression, the \peii\ decay branching
ratio is highly susceptible to small non-(\VmA) contributions from new
physics, making this decay a particularly suitable subject of study, as
discussed in, e.g., Refs.~\cite{Shr81,Sha82,Loi04,Ram07,Cam05,Cam08}.
This sensitivity provides the primary motivation for the ongoing
PEN\cite{PENweb} and PiENu\cite{PiENuWeb} experiments.  Of all the
possible ``new physics'' contributions in the Lagrangian, \peii\ is
directly sensitive to the pseudoscalar one, while other types enter
through loop diagrams.  At the precision of $10^{-3}$, $R_{e/\mu}^\pi$
probes the pseudoscalar and axial vector mass scales up to 1,000\,TeV
and 20\,TeV, respectively\cite{Cam05,Cam08}.  For comparison, unitarity
tests of the Cabibbo-Kobayashi-Maskawa (CKM) matrix and precise
measurements of several superallowed nuclear beta decays constrain the
non-SM vector contributions to $>20\,$TeV, and scalar ones to $>10\,$TeV
\cite{PDG18}.  Although scalar interactions do not directly contribute
to $R_{e/\mu}^\pi$, they can do so through loop diagrams, resulting in a
sensitivity to new scalar interactions up to 60\,TeV \cite{Cam05,Cam08}.
The subject was recently reviewed in Refs.~\cite{Bry11,Poc14}.  In
addition, $R_{e/\mu}^{\pi,\,\text{exp}}$ provides limits on the masses of
certain SUSY partners\cite{Ram07}, and on anomalies in the neutrino
sector\cite{Loi04}.  Recent intriguing indications of LU violation in
B-meson decays make the subject additionally interesting (for a recent
review see \cite{Cie17}).

\section{The PEN experiment}

\begin{figure}[tb]
  \centerline{
    \includegraphics[width=0.9\linewidth]{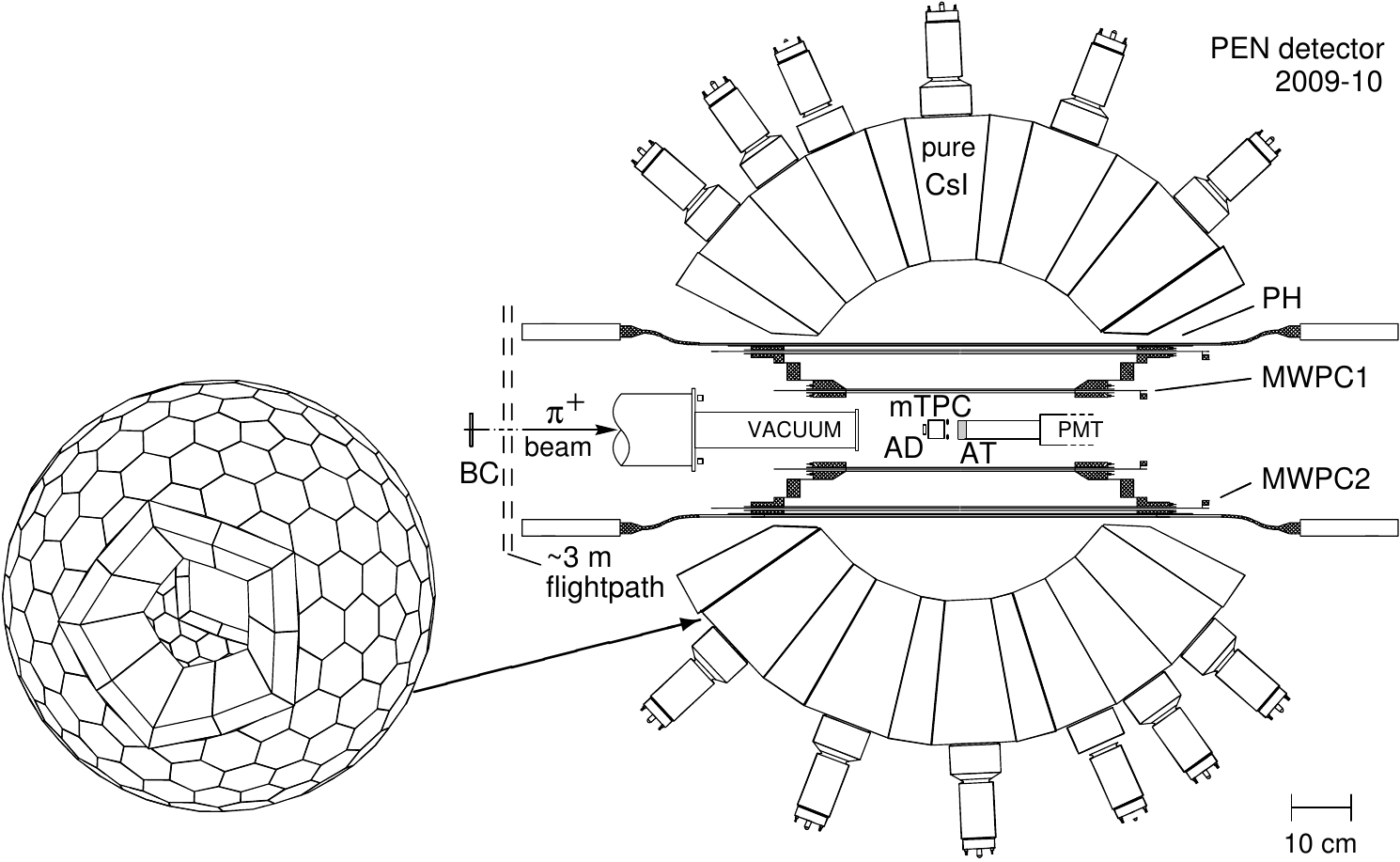}
             }
    \caption{Schematic cross section of the PEN apparatus, shown in the
      2009-10 running configuration.  See text for explanation of
      abbreviations, and \cite{Frl04a} for details concerning the
      detector performance.}
      \label{fig:PEN_det}
\end{figure}

PEN is a measurement of $R_{e/\mu}^{\pi}$ carried out in three runs in
2008--2010 at the Paul Scherrer Institute (PSI) by a collaboration of
seven US and European institutions\cite{PENweb}, with the aim to reach
\begin{equation}
 %  \frac
      {\Delta R_{e/\mu}^{\pi}}/{R_{e/\mu}^{\pi}}
       \simeq 5 \times 10^{-4} \,.    \label{eq:pen_goal}
\end{equation}
The PEN experiment uses the key components of the PIBETA apparatus with
additions and modifications suitable for a dedicated study of the
\peii\ and \peiig\ decay processes.  The PIBETA detector has been
described in detail in \cite{Frl04a}, and used in a series of
measurements of rare allowed pion and muon decay channels
\cite{Poc04,Frl04b,Byc09,Poc14}.  The major component of the PEN
apparatus, shown in Figure~\ref{fig:PEN_det},
is the spherical large-acceptance ($\sim\,3\pi$\,sr) electromagnetic
shower calorimeter.  The calorimeter consists of 240 truncated hexagonal
and pentagonal pyramids of pure CsI, 22\,cm or 12 radiation lengths
deep.  The inner and outer diameters of the sphere are 52\,cm and
96\,cm, respectively.  Beam particles entering the apparatus with
$p\simeq 75$\,MeV/$c$ are first tagged in a thin upstream beam counter
(BC) and refocused by a triplet of quadrupole magnets.  Following a
$\sim 3$\,m long flight path they pass through a 5\,mm thick active
degrader (AD) and a low-mass mini time projection chamber (mTPC), to
reach a 15\,mm thick active target (AT) where the beam pions stop and
\textbf{decay at rest}.  Decay particles are tracked non-magnetically in
a pair of concentric cylindrical multiwire proportional chambers
(MWPC1,2) and an array of twenty 4\,mm thick plastic hodoscope detectors
(PH), all surrounding the active target.  The BC, AD, AT and PH
detectors are all made of fast plastic scintillator material and read
out by fast photomultiplier tubes (PMTs).  Signals from the beam
detectors are sent to waveform digitizers, running at 2\,GS/s for BC,
AD, and AT, and at 250\,MS/s for the mTPC.

Measurements of pion decay at rest, as in the PEN experiment, must deal
with the challenge of separating the $\pi\to e\nu$ and $\pi\to\mu\to e$
events with great confidence.  Hence, a key source of systematic
uncertainty in PEN is the hard to measure low energy tail of the
detector response function.  The tail is caused by electromagnetic
shower leakage from the calorimeter, mostly in the form of photons.  In
addition, if not properly identified and accounted for, other physical
processes can contribute events to the low energy part of the spectrum,
as well as higher energy events above the muon beta decay (``Michel'')
endpoint.  One such process is the ordinary pion decay into a muon in
flight, before the pion is stopped, with the resulting muon decaying
within the time gate accepted in the measurement.  Another is the
unavoidable physical process of radiative decay.  The latter is measured
and properly accounted for in the PEN apparatus, as was demonstrated in
PIBETA analyses \cite{Byc09}.  Shower leakage and pion decays in flight
can be appropriately characterized only if the $\pi\to\mu\to e$ chain
can be well separated from the direct $\pi \to e$ decay in the target.
Methods used by PEN to separate the two decay paths are discussed in
\cite{Poc14,Poc17} and references therein.

\begin{figure}[t]
  \parbox{0.65\linewidth}{
    \centerline{\includegraphics[width=0.95\linewidth]{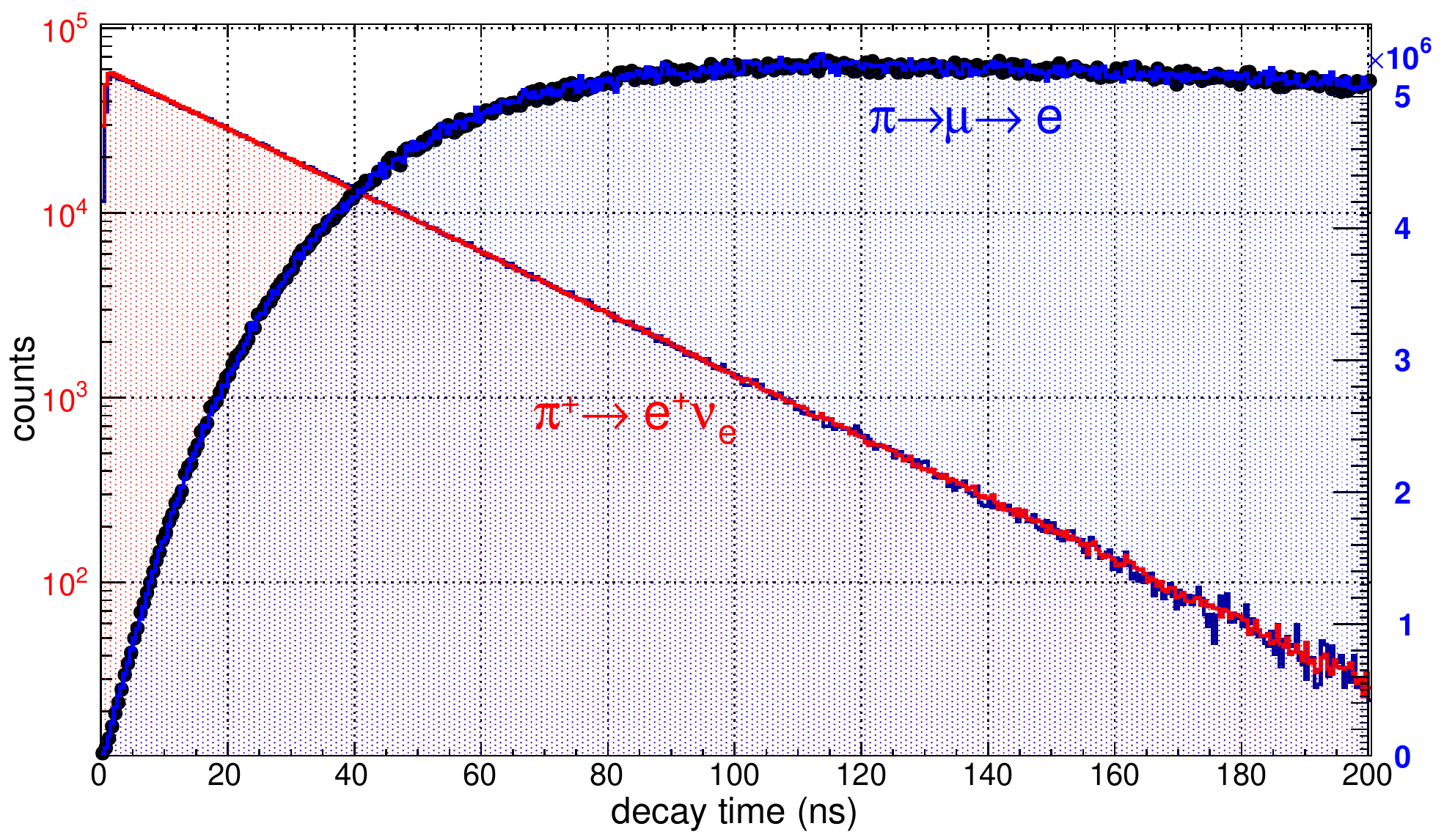}}
%%  \centerline{\includegraphics[width=0.9\linewidth]{unity_pi2e_10} }
%%  \centerline{\includegraphics[width=0.9\linewidth]{unity_michel_10} }
                         }
  \hspace*{\fill}
  \parbox{0.33\linewidth}{\vspace*{-12pt}
    \caption{Data points: decay time spectra for the clean samples of
      $\pi\to e\nu$ (red) and $\pi\to\mu\to e$ (blue) events from the
      2010 PEN data set.  Curves: PEN Geant4 simulation of
      the same processes.  Agreement demonstrates clean separation of
      the two processes. \label{fig:decaytim} }}
\end{figure}

\begin{figure}[b]
  \parbox[t]{0.65\linewidth}{\hrule height 0pt width 0pt
    \centerline{\includegraphics[width=\linewidth]{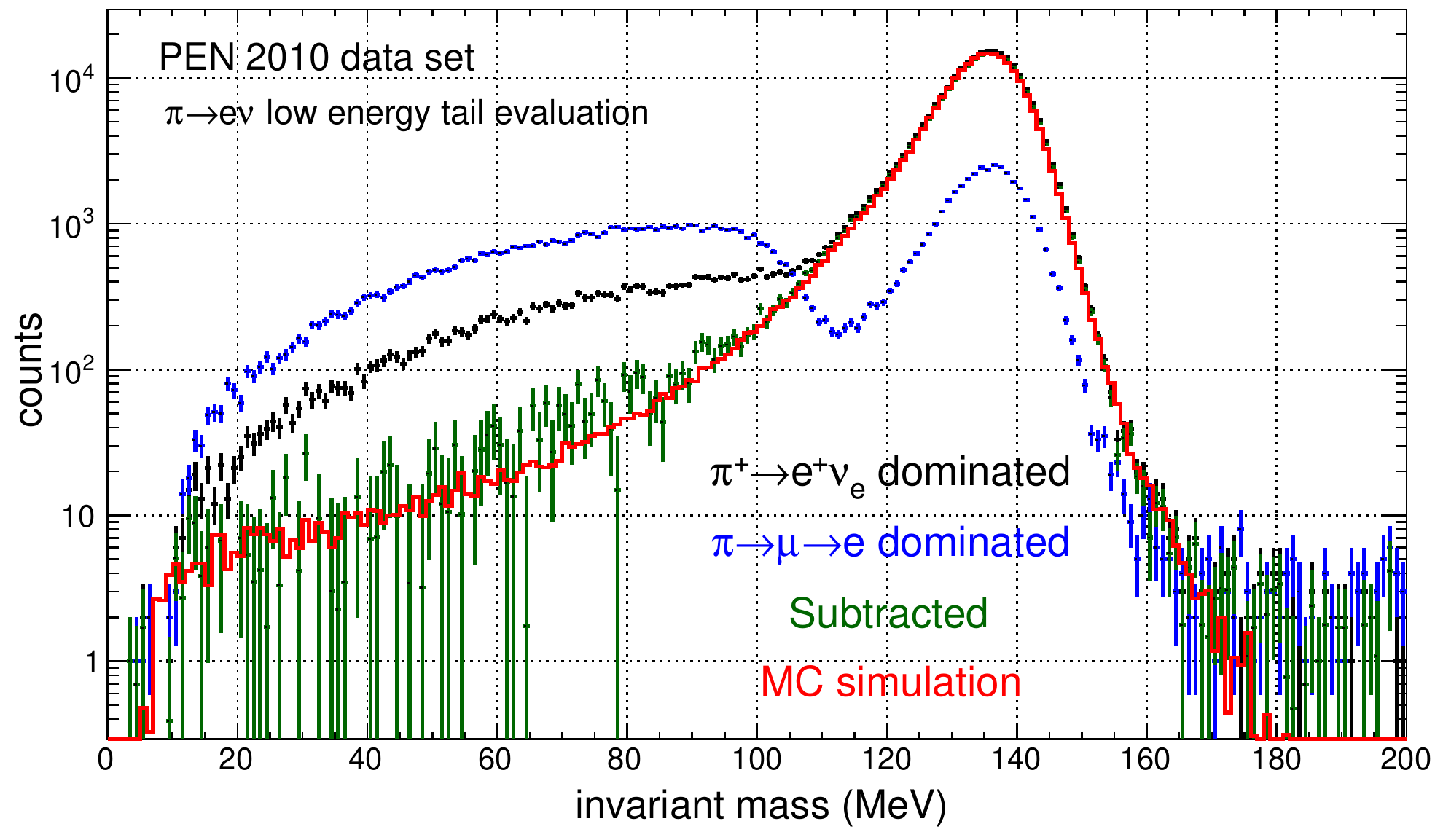}}
                         }
  \parbox[t]{0.33\linewidth}{\vspace*{-8pt}
    \caption{Low energy tail of the calorimeter response (green),
      evaluated as a difference between the $\pi$$\to$e$\nu$ (black) and
      $\pi$$\to$$\mu$$\to$$e$ (red) dominated events, collected in 2010
      using a dedicated ``tail'' trigger, designed to maximize
      $\pi$$\to$$e$$\nu$ over $\pi$$\to$$\mu$$\to$$e$
      yield. \label{fig:tail}
            }             }
\end{figure}

In all, PEN has accumulated well over $2\times 10^7$ raw $\pi\to e\nu$
events and well over $10^8$ raw $\pi\to\mu\to e$ events, allowing for
generous data quality selections.  The quality of PEN data is best
illustrated in figures~\ref{fig:decaytim} and \ref{fig:tail}
which show the representative decay time (time following $\pi$ stop in
AT) and energy spectra for the two classes of events.%% (2010 data).

\noindent
The experimental branching ratio $R_{e/\mu}^{\pi,\,\text{exp}}$ is determined
as follows:
\begin{align}
  R_{e/\mu}^{\pi,\,\text{exp}}
       & = \frac{N_{\pi\to e\nu}^{\text{peak}}
                (1+\epsilon_\text{tail})}  {N_{\pi\to \mu\nu}}
        \cdot \frac{f_{\pi\to \mu\to \text{e}}(T_\text{e})}
                {f_{\pi\to \text{e}\nu}(T_\text{e})}
         \cdot \frac{\epsilon(E_{\mu\to e\nu\bar{\nu}})_\text{\tiny MWPC}}
                {\epsilon(E_{\pi\to e\nu})_\text{\tiny MWPC}}
         \cdot \frac{A_{\pi\to\mu\to e}}{A_{\pi\to e\nu}}  \\
         & = \frac{N_{\pi\to e\nu}^{\text{peak}}}
                {N_{\pi\to \mu\nu}} \cdot (1+\epsilon_\text{tail})
             \cdot r_f \cdot r_{\epsilon} \cdot r_A\,,
 \label{eq:BR_exp_PEN}   
\end{align}
where $\epsilon_\text{tail}$ is the low energy tail fraction of the
$\pi\to e\nu$ response buried under the $\pi\to\mu\to e$ signal, $r_f$
is the ratio of the decay fractions for the two processes in the
observed decay time gates, $r_{\epsilon}$ is the ratio of the MWPC
efficiency for the two processes (not $\equiv 1$ because of the positron
energy dependence of the energy deposited in chamber gas), and $r_A$ is
the ratio of the geometrical acceptances for the two processes,
evaluated from simulation.  For the event selection criteria used in the
evaluation of $R_{e/\mu}^\pi$, $r_A$ is practically indistinguishable
from unity.  The magnitudes and associated uncertainties of the
quantities needed to determine $R_{e/\mu}^\pi$, and given in
equation~\ref{eq:BR_exp_PEN}, are summarized in table~\ref{tab:uncerts}.
\begin{table}[t]
 \begin{center}

  \begin{tabular}{llcc}
   \Hline \\[-10pt]
    Type & Observable & Value & $\Delta R_{e/\mu}^{\pi}/R_{e/\mu}^{\pi}$ \\
   \hline \\[-10pt]
    {Systematic:}
      & {$\Delta \epsilon_{\text{tail}}$}
      & $\simeq 0.025$
      & $\begin{cases}
          \simeq  0.001^\text{exp}  \\
          2\times 10^{-4}|^\text{\tiny MC}_{\text{\tiny goal}}
         \end{cases}$ \\
      & &  &            \\[-9pt]
    & {$r_f$} & 0.046 & $1.8\times 10^{-4}$\\
    & {$r_\epsilon$} & $\simeq .99$ & $< 10^{-4}$ \\
    & $r_A$ & $\simeq 1$ & $\leq 10^{-4}$\\
    & $N_{\pi_\text{DIF} \to \text{e}\nu}/N_{\pi\to \text{e}\nu}$
          & $ < 2\times 10^{-3}$ & $10^{-6}- 10^{-5}$\\ 
    & $N_{\pi_\text{DIF}\to \mu\nu}/N_{\pi\to \mu\nu}$
          & $ 2.3\times 10^{-3}$&$10^{-6}-10^{-5}$\\
    & $N_{\mu_\text{DIF}\to \text{e}\nu\bar{\nu}}/N_{\mu\to \nu\bar{\nu}}$
          & $1.4\times 10^{-4}$ & $10^{-6}-10^{-5}$\\
   \hline \\[-10pt]
    {Statistical: }
    &  $\Delta N_{\pi\to \text{e}\nu}/N_{\pi\to \text{e}\nu}$
         & & $\simeq   2.9\times 10^{-4}$\\ 
   \hline \\[-10pt]
    {Overall} & goal  & & $5\times 10^{-4}$\\
   \Hline
  \end{tabular}
  \caption{Uncertainty budget for the determination of the $\pi\to
    e\nu(\gamma)$ branching ratio in PEN, including the dominant
    sources of systematic and statistical uncertainties.  Label ``DIF'' 
    denotes decay in flight of the particle so marked.}
  \label{tab:uncerts}

 \end{center}
\end{table}
The systematic and statistical uncertainties are comparable in magnitude.

Perhaps the toughest nut to crack among the leading systematic
uncertainties relates to $\epsilon_{\text{tail}}$, the infamous low
energy tail correction to the branching ratio in
equation~\ref{eq:BR_exp_PEN}, caused by shower leakage outside the CsI
electromagnetic calorimeter.  To begin with, the low energy ``tail''
cannot be measured with sufficient accuracy concurrently with the
branching ratio measurement, due to the $\sim$\,5 orders of magnitude
higher background of Michel positrons emanating from the $\pi\to\mu\to
e$ decay chain.  This is illustrated in figure~\ref{fig:tail} which
shows the experimentally determined low energy ``tail'' evaluated
through subtraction of a Michel background dominated event sample from a
$\pi\to e\nu$ dominated event sample.  The events shown were collected
during the 2010 run by means of a specially constructed ``tail''
trigger, designed to suppress the yield of Michel background events.
Thus obtained experimental ``tail'' is in excellent agreement with the
Monte Carlo simulated PEN detector response, as seen in the figure.
However, the last factor of 3--5 in precision must be provided by the
simulation, because it is not available from the data.

\begin{figure}[t]
  \centerline{
    \parbox{0.9\linewidth}{
      \includegraphics[width=\linewidth]{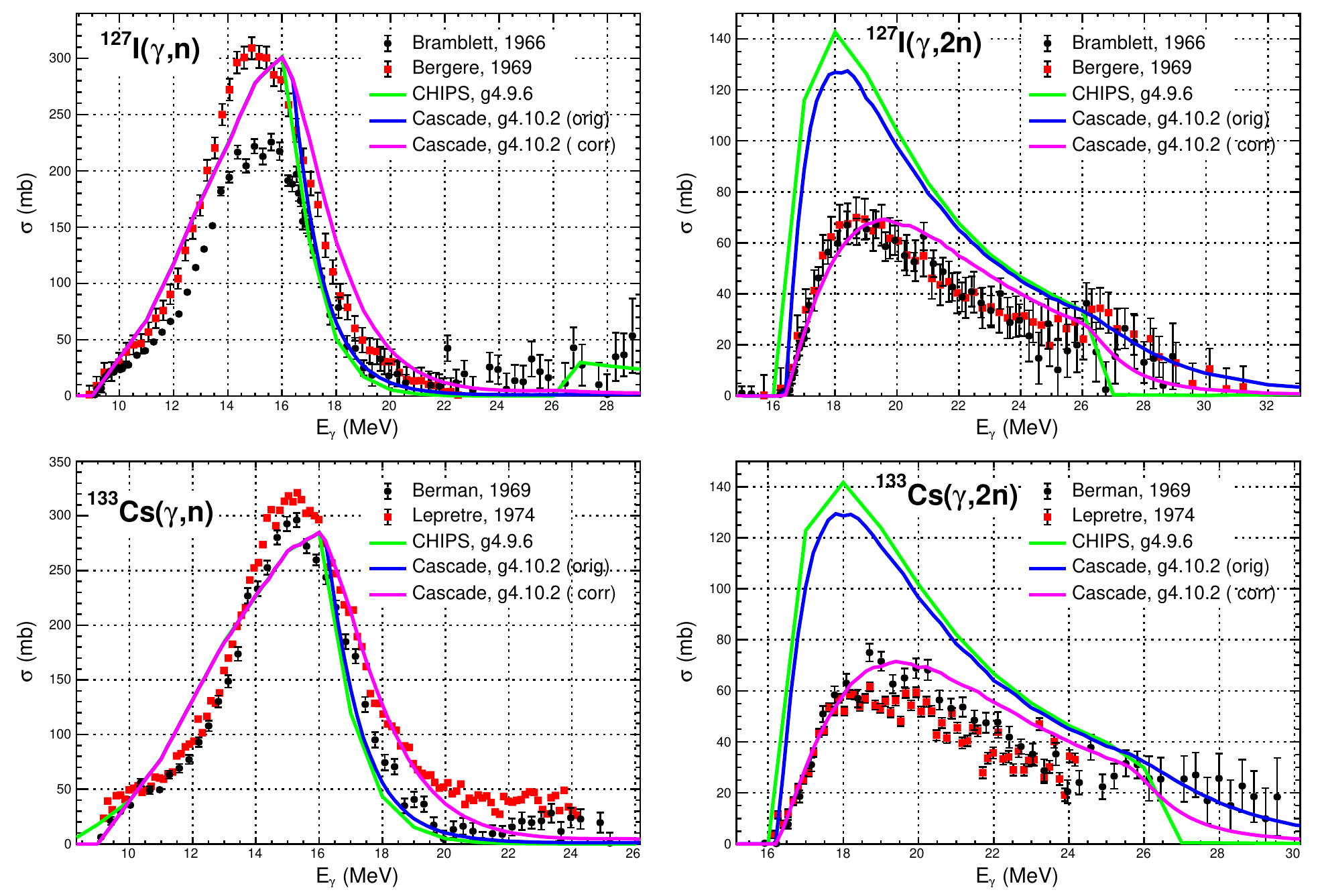}
             }             }
  \hspace*{\fill}
  \caption{\small Data points and curves: experimental and cascade model
    calculated photonuclear cross sections, respectively, used as input
    for calculations of the PEN CsI calorimeter low energy response tail.
    Further model improvements of both $(\gamma,n)$ and
    $(\gamma,2n)$ are under way.}
   \label{fig:photonuc}
\end{figure}

The principal complication in simulating the PEN low energy tail
response stems from the presence of photonuclear reactions in the
calorimeter material.  The greatest distortion of the purely
electromagnetic shower response comes about when an energetic shower
photon is absorbed by a Cs or I nucleus with subsequent emission of one
or more neutrons.  The emitted neutrons, in turn, have a significant
probability of escaping the calorimeter volume without depositing their
full energy in it, and thus shifting ``peak'' energy events down into
the low energy ``tail.''  Unfortunately, the situation with the relevant
$(\gamma,n)$ and $(\gamma,2n)$ cross sections is not satisfactory on two
counts, as illustrated in figure~\ref{fig:photonuc} and discussed below.
First, the $(\gamma,n)$ cross section data sets, especially that for
$^{127}\text{I}$, are not in full internal agreement.  Second, the
various cascade model calculations, especially for the $(\gamma,2n)$
cross sections, used in the Geant4 Monte Carlo package have been in
dramatic disagreement with measured values (by about a factor of 2), and
have changed significantly from one Geant4 release version to another.
After much interaction with the authors and maintainers of Geant, we
have found ways to modify and correct the cascade model calculations of
$(\gamma,2n)$ cross sections, as shown in figure~\ref{fig:photonuc}.  As
of this writing, members of the PEN collaboration are actively working
on bringing the model calculations of photonuclear cross sections fully
in line with measured data, and on quantifying the impact of the
existing ambiguities in the $(\gamma,n)$ cross sections on the PEN
$\epsilon_{\text{tail}}$ correction.  Once this task is completed, the
branching ratio will be ready for evaluation.

\section{Summary}

During three production runs, in 2008, 2009 and 2010, the PEN experiment
recorded more than $2 \times 10^7$ $\pi\to e\nu$, and over $10^8$
$\pi\to\mu\to e$ events, including significant numbers of pion and muon
hard radiative decay events.  A comprehensive blinded analysis to
extract a new experimental value of $R_{e/\mu}^{\pi}$ is nearing
completion, with expected precision of $\Delta R/R < 10^{-3}$.  PiENu, a
similar experiment at TRIUMF \cite{PiENuWeb} with significantly
different systematics has a similar precision goal.  The near term
future will thus bring about a substantial improvement in the limits on
$e$-$\mu$ lepton universality, and in the related limits on non-SM,
non-(\VmA) processes and couplings.

In addition, new results will be forthcoming in the analysis, not
discussed here in detail, of radiative $\pi^+\to e^+\nu_e\gamma$ and
$\mu^+\to e^+\nu_{e}\bar{\nu}_{\mu}\gamma$ decays in the PEN data set.
The former will bring about new constraints on the pion weak form
factors, the $SD^-$ term $\propto (F_V-F_A)^2$, with attendant
improvement of the chiral low-energy constants.  The muon radiative
decay analysis will sharpen existing constraints on the $\bar{\eta}$
Michel parameter, and on non-(\VmA) weak interaction couplings
\cite{Van06,Mun12}.

However, even subsequent to the completion of the current PEN and PiENu
data analyses, there will remain considerable room for improvement of
experimental precision of $R_{e/\mu}^{\pi}$, with high payoff in terms of
limits on physics not included in the present standard model.  This work
remains relevant and complementary to direct searches on the energy
frontier, underway at particle colliders, providing valuable theoretical
model cross checks.

%%  \Acknowledgements
\medskip
\centerline{\bfseries ACKNOWLEDGEMENTS}
\smallskip

The PEN collaboration has received support from the US National Science
Foundation, the Paul Scherrer Institute, and the Russian Foundation for
Basic Research.


\begin{thebibliography}{99}

%%
%%  bibliographic items can be constructed using the LaTeX format in SPIRES:
%%    see    http://www.slac.stanford.edu/spires/hep/latex.html
%%  SPIRES will also supply the CITATION line information; please include it.
%%

 \bibitem{Fey58} R.P.~Feynman and M.~Gell-Mann,
   %%   \emph{Theory of Fermi interaction},
   \PR \textbf{109} (1958) 193.

 \bibitem{Faz59} T.~Fazzini et al.,
   %%  G.~Fidecaro, A.W.~Merrison, H.~Paul and A.V.~Tollestrup,
   %%  \emph{Electron decay of the pion},
   \PRL \textbf{1} (1959) 247. 

 \bibitem{Ber58} S.M.~Berman,
   %%  \emph{Radiative corrections to pion $\beta$ decay},
   \PRL \textbf{1} (1958) 468. 

 \bibitem{Kin59} T.~Kinoshita,
   %%  \emph{Radiative corrections to $pi$-e decay},
   \PRL \textbf{2} (1959) 477.

 \bibitem{And60} H.L.~Anderson et al.,
  %%  T.~Fujii, R.H.~Miller and L.~Tau,
  %%  \emph{Branching ratio of the electronic mode of positive pion decay},
  \PR \textbf{119} (1960) 2050.

\bibitem{DiC64} E.~Di~Capua et al.,
  %%  R.~Garland, L.~Pondrom and A.~Strelzoff,
  %%  \emph{Study of the decay $\pi \to e+\nu$},
  \PR \textbf{133} (1964) B1333.

\bibitem{Bry82}
  %% For a detailed discussion of the $\pi_{e2(\gamma)}$ decays see,
  %% e.g.,
  D.A.~Bryman, P.~Depom\-mier and C.~Leroy,
  %%  \emph{$\pi\to e\nu$, $\pi\to e\nu\gamma$ decays and related processes},
  Phys.\ Rep.\ \textbf{88} (1982) 151.

\bibitem{Mar93} W.J.~Marciano and A.~Sirlin,
  %%  \emph{Radiative corrections to $\pi_{l2}$ decays},
  \PRL \textbf{71} (1993) 3629.

\bibitem{Fin96} M.~Finkemeier,
  %%  \emph{Radiative corrections to $\pi_{l2}$ and $K_{l2}$ decays},
  Phys.\ Lett.\ B \textbf{387} (1996) 391.
  %%   [hep-ph/9505434].

\bibitem{Cir07} V.~Cirigliano and I.~Rosell,
  %%  \emph{ Two-loop effective theory analysis of $\pi(K)\to e
  %%  \bar{\nu}_e [\gamma]$ branching ratios}, 
  \PRL \textbf{99} (2007) 231801.
  %%   [hep-ph/0707.3439].

\bibitem{Bri92a} D.I.~Britton et al.,
  %%  \emph{Measurement of the $\pi^+ \to e^+ \nu$ branching ratio},
  \PRL \textbf{68} (1992) 3000.

\bibitem{Bri92b}  D.I. Britton et al., 
  %%  \emph{Improved search for massive neutrinos in $\pi^+ \to e^+ \nu$ decay},
  \PR\ D \textbf{46} (1992) R885.

\bibitem{Cza93} G.~Czapek et al.,
  %%   \emph{Branching ratio for the rare pion decay into e$^+$ and $\nu_e$},
   \PRL \textbf{70} (a993) 17.

\bibitem{Agu15} A. Aguilar-Arevalo et al., 
  %%  \emph{Improved measurement of the $\pi\to e\nu$ branching ratio}, 
  \PRL \textbf{115} (2015) 071801.
  %%  [hep-ex/1506.05845].

\bibitem{Shr81} R.E.~Shrock,
  %%  \emph{General theory of weak processes involving neutrinos},
  \PR\ D \textbf{24} (1981) 1232.

\bibitem{Sha82} O.U.~Shanker,
  %%  \emph{$\pi_{\ell2}$, $K_{\ell3}$ and $K^0$--$\bar{K}^0$
  %%  constraints on leptoquarks and supersymmetric particles},
  Nucl.\ Phys.\ \textbf{B204} (1982) 375.

\bibitem{Loi04} W.~Loinaz  et al.,
  %%  N.~Okamura, S.~Rayyan, T.~Takeuchi, L.C.R.~Wijewardhana,
  %%  \emph{NuTeV anomaly, lepton universality, and nonuniversal
  %%  neutrino-gauge couplings},
  PR\ D \textbf{70} (2004) 113004.
  %%  [hep-ph/0403306].

\bibitem{Ram07} M.J.~Ramsey-Musolf, S.~Su and S.~Tulin,
  %%  \emph{Pion leptonic decays and supersymmetry},
  \PR\ D \textbf{76} (2007) 095017.
  %%  [hep-ph/0705.0028].

% \bibitem{Dav94} S.~Davidson, D.~Bailey, B.~A.~Campbell, Z.\ Phys.\ C
%   \textbf{61} (1 994) 613-644. 

\bibitem{Cam05} B.A.~Campbell and D.W.~Maybury,
  %%  \emph{Constraints on scalar couplings from $\pi^{\pm}\to
  %%  \ell^{\pm} \nu_{\ell}$}, 
  Nucl.\ Phys. \textbf{B709} (2005) 419.
  %%  [hep-ph/0303046].

\bibitem{Cam08} B.A.~Campbell and A.~Ismail,
  %%  \emph{Leptonic pion decay and physics beyond the electroweak
  %%  standard model}, %\emph{e-Print} 
  hep-ph/0810.4918.

\bibitem{PENweb} {\href{http://pibeta.phys.virginia.edu}{
                \texttt{http://pibeta.phys.virginia.edu}}},
                 and links therein.

\bibitem{PiENuWeb} {\href{http://pienu.triumf.ca}{
                \texttt{http://pienu.triumf.ca}}},
                and links therein.

%%%  came to here  %%%
                
%% \bibitem{PDG16} K.A. Olive \etal\ (Particle Data Group),
  %%  \emph{Review of Particle Physics},
%  \emph{Chin.\ Phys.\ C} \textbf{38} (2014) 090001, and 2015 update. 
%%   \emph{Chin.\ Phys.\ C} \textbf{40} (2016) 100001.

 \bibitem{PDG18} M. Tanabashi et al. (Particle Data Group),
   Phys.\ Rev.\ D \textbf{98} (2018) 030001. 
   
 \bibitem{Bry11} D.A.~Bryman et al.,
   %%  W.J.~Marciano, R.~Tschirhart and T.~Yamanaka,
   %%  \emph{Rare kaon and pion decays: Incisive probes for new physics
   %%  beyond  the standard model},  
   Annu.\ Rev.\ Nucl.\ Part.\ Sci. \textbf{61} (2011) 331.

 \bibitem{Poc14} D.~Po\v{c}ani\'c,  E.~Frle\v{z}, A. van der Schaaf,
   J. Phys.\ G: Nucl.\ Part.\ Phys.\ \textbf{41} G41 (2014 114003).
   
 \bibitem{Cie17} G.~Ciezarek et al.,
   %% M.F.~Sevilla, B.~Hamilton, R.~Kowalewski, T.~Kuhr, V.~L\"uth, and
   %% Y.~Sato,
   Nature \textbf{546} (2017) 228.
   %% [DOI:10.1038/nature22346]

 \bibitem{Frl04a} E.~Frle\v{z}, D. Po\v{c}ani\'c, et al.,
   %%  \emph{Design, commissioning and performance of the PIBETA detector
   %%  at PSI}, 
   \NIM\ A \textbf{526} (2004) 300.
   %%  [hep-ex/0312017].

 \bibitem{Poc04} D.~Po\v{c}ani\'c et al.,
   %%  \emph{Precise measurement of the $\pi^+ \to \pi^0 e^+ \nu$
   %%  branching ratio}, 
   \PRL\ \textbf{93} (2004) 181803.
   %%  [hep-ex/0312030].

\bibitem{Frl04b} E.~Frle\v{z} et al.,
  %%  \emph{Precise measurement of the pion axial form-factor in the
  %%  $\pi^+ \to e^+ \nu \gamma$ decay}, 
  \PRL\ \textbf{93} (2004) 181804.

\bibitem{Byc09} M.~Bychkov et al.,
  %%  \emph{New precise measurement of the pion weak form factors in
  %%  $\pi^+ \to e^+ \nu \gamma$ decay},  
  \PRL\ \textbf{103} (2009) 051802.
  %%  [hep-ex/0804.1815].

\bibitem{Poc17} D.~Po\v{c}ani\'c et al., 
   Proceedings of Science, PoS HQL2016 (2017) 042.

\bibitem{Van06} B.A.~VanDevender, PhD thesis, University of Virginia (2006).

\bibitem{Mun12} E.~Munyangabe, PhD thesis, University of Virginia (2012).

\end{thebibliography}
\end{document}